\documentclass[intlimits,twoside,a4paper]{article}

\usepackage[cp1251]{inputenc}

\usepackage[eqsecnum]{cmpj3}

\usepackage{bm}

\usepackage{comment}

\issue{2024}{27}{3}{33601}
\doinumber{10.5488/CMP.27.33601}

\title{Partition function zeros of zeta-urns}
\author[P. Bialas, Z. Burda, D. A. Johnston]{P. Bialas\orcid{0000-0003-0704-9168}\refaddr{label1}, Z. Burda\orcid{0000-0002-9656-9570}\refaddr{label2}, D. A. Johnston\orcid{0000-0003-0556-3200}\refaddr{label3}\thanks{Corresponding author: \email{D.A.Johnston@hw.ac.uk}.}}

\addresses{
\addr{label1}Institute of Applied Computer Science, Jagiellonian University, ul. Lojasiewicza 11, 30-348 Krak\'ow, Poland
\addr{label2}AGH University of Krakow, Faculty of Physics and Applied Computer Science, 
al. Mickiewicza 30, 30-059 Krak\'ow, Poland
\addr{label3}School of Mathematical and Computer Sciences, Heriot-Watt University, Riccarton, Edinburgh EH14 4AS, UK
}

\Keywords{Lee-Yang and Fisher zeroes, critical exponents, first order phase transitions, second order phase transitions}

\date{Received November 30, 2023, in final form January 24, 2024}
\begin{document}

\maketitle

\begin{abstract}
We discuss the distribution of partition function  zeros for the grand-canonical ensemble 
of the zeta-urn model, where tuning a single parameter can give a first or any higher order condensation transition. We compute the locus of zeros for finite-size systems and test scaling 
relations describing the accumulation of zeros near the critical point against theoretical predictions for both the first and higher order transition regimes.

\printkeywords
\end{abstract}

\section{Introduction}
In this paper we highlight the fact that a simple model of weighted partitions of indistinguishable particles into boxes, the zeta-urn model \cite{bbj,balls2, dgc} in a grand-canonical ensemble where the number of {\em boxes} can fluctuate \cite{bbj2,g}, provides a useful illustrative example for exploring the finite size scaling of partition function zeros, since its finite size partition function is, at least in principle,  exactly calculable and the order of its phase transition (condensation) may be tuned by varying a single parameter. The thermodynamic limit is also amenable to evaluation via saddle point methods.

In section \ref{sec:zeta-urn} we first describe the formulation of the zeta-urn model in the canonical ensemble, since the partition function of the grand-canonical ensemble builds on the canonical partition function. The presence of a transition in the model is revealed by the breakdown of the saddle point approximation in the thermodynamic limit and its nature, a real space condensation \cite{e,gss,eh,gl,wbbj,mez,emz,emz2,j,bb,bbw2,g1}, can be elucidated by explicitly evaluating the  fraction of sites with $q$ particles in a given configuration in finite size systems as the mean particle density is varied.   We then discuss the thermodynamic limit in the grand-canonical ensemble and note that first or higher order transitions may be observed by tuning the power that appears in the weights for single box configurations. 
In section \ref{sec:PFZ} we review some general features of partition function zeros~\cite{LY1,LY2,Fi64,BDL,Abe1,Abe2,Abe3, Abe4,SuzukiLY1,SuzukiLY2,SuzukiLY3,SuzukiLY4} and their finite size scaling for phase transitions of different orders \cite{wjk}, using both the density of zeros at a given system size and the scaling of the position of a given zero as the system size varies.
In sections \ref{sec:FSS}, \ref{sec:RES} we investigate the application of these methods to studying the partition function zeros of  the grand-canonical zeta-urn model.  For $\beta>1$, the model displays a phase transition and we discuss the first order and second/higher order regimes in subsections \ref{first} and \ref{second}, respectively. The locus of zeros still displays interesting properties for values of $\beta$ for which there is no phase transition (i.e., $\beta \leqslant 1$, including negative values) and we outline these in subsection \ref{unphysical}. 
We close with a summary and brief discussion in section \ref{sec:summary}.

\section{Zeta-urn models}
\label{sec:zeta-urn}

In the canonical ensemble, the partition function of the zeta-urn model, which is a particular case of a balls-in-boxes model,  is defined by 
\cite{bbj}
\begin{equation}
    Z_{S,N} = \sum_{(s_1,\ldots,s_N)} 
    w(s_1) \ldots w(s_N) \delta_{S- (s_1+\ldots +s_N)},
    \label{eq:ZSN}
\end{equation}
with $w(s) = s^{-\beta}$ \cite{bbj}. The name comes from the presence of the Riemann zeta function in the overall normalization and in the saddle point equations discussed below.
The model describes power-law weighted distributions of $S$ indistinguishable particles into $N$ boxes. 
The $s_i$'s denote the occupation numbers of the boxes $i=1,\ldots,N$ and we take $s_i \geqslant 1$.
The lowercase $\delta$ is the Kronecker delta.

A useful relation for evaluating $Z_{S,N}$ exactly follows from considering configurations with $q$ balls in a given box. In that case the remaining $N-1$ boxes contain $S-q$ balls and are thus described by the partition function $Z_{S-q,N-1}$. This gives the recurrence relation
\begin{equation}
    Z_{S,N}=\sum_{q=1}^{S-N+1}w(q) Z_{S-q,N-1}
\label{eq:Z-recurrence}    
\end{equation}
for $S\geqslant N\geqslant q$, and $Z_{S,1}=w(S)$. 
The thermodynamic limit may be taken with both $N$ and $S$ being sent to infinity in 
a fixed ratio, i.e., $S\rightarrow \infty , \  \frac{N}{S} \rightarrow r$,  
where $r \in (0,1)$ is
the reciprocal of the average particle density $\rho\equiv\frac{S}{N} =\frac{1}{r}$.
The behaviour in the thermodynamic limit is encapsulated in a thermodynamic potential (a ``free energy density'' for the system)
\begin{equation}
 \phi(r) = \lim_{S \rightarrow \infty} \frac{1}{S} \ln Z_{S,N} 
\label{phi}
\end{equation}
at fixed $r$, that is, 
$\frac{N}{S} \rightarrow r \in (0,1)$. 
The potential $\phi(r)$  gives the rate of the asymptotic growth of the partition function as $S\to \infty$ at fixed $r$: $Z_{S,N} \propto \re^{S \phi(r)}$ and can be calculated using the saddle point  method.
The saddle point solution only holds
for $r>r_{\rm c}$ 
\begin{equation}
    r_{\rm c} = \frac{1}{\rho_c}=  \frac{\zeta(\beta)}{\zeta(\beta-1)} .
    \label{rcr}
\end{equation}
To understand what is happening for $r<r_c$ (i.e., $\rho >\rho_c$), it is useful to look at the fraction of sites with $q$ particles in a given configuration:
\begin{equation}
\pi(q) = \frac{1}{N} \sum_{i=1}^N \delta_{q-s_i} \; ,
\label{pi_definition}
\end{equation}
whose ensemble average is given by
\begin{equation}
    \langle \pi(q) \rangle_{S,N} = \frac{w(q) Z_{S-q,N-1}}{Z_{S,N}} \; .
    \label{piq}
\end{equation} 
For $r<r_c$ ($\rho>\rho_c$), we can still use (\ref{eq:Z-recurrence})  to evaluate 
(\ref{piq}) exactly, even though the saddle point evaluation has broken down. If $\rho>\rho_c$, 
a peak appears in $\langle \pi(q) \rangle_{S,N}$ whose maximum is at $S(\rho - \rho_c )$ and whose sharpness increases with an increasing system size.
This peak is a signal that an extensive fraction of the particles has condensed into a single box, which is a real space condensation \cite{gss,gl,fls,cg1,agl,cg2,jcg,g3}. For $\rho<\rho_c$, on the other hand, no box is distinguished and this phase is called ``fluid''.

It is possible to consider (amongst others)  an  ensemble
with a variable number of boxes $N$ controlled by a
chemical potential $\mu$.  The corresponding partition function in this case is
\begin{equation}
    Z_{S,\mu} = \sum_{N=1}^S \re^{-\mu N} Z_{S,N},
    \label{ZSmu}
\end{equation}
where $Z_{S,N}$ is the canonical partition function defined in (\ref{eq:ZSN}).
With a slightly non-standard terminology, it has been customary to call (\ref{ZSmu})
the ``grand canonical'' partition function \cite{bbj2}.

In the thermodynamic limit, $S\rightarrow \infty$,
the corresponding grand-canonical thermodynamic potential is defined as 
\begin{equation}
 \psi(\mu) = \lim_{S\rightarrow \infty} \frac{1}{S} 
 \ln Z_{S,\mu} .
\label{eq:psi}
\end{equation}
$\psi(\mu)$ is related to 
$\phi(r)$ by a Legendre-Fenchel transform
\begin{equation}
    \psi(\mu) = 
    \sup_{r \in [0,1]} \left(-\mu r + \phi(r)\right) 
    \label{psi_phi} .
\end{equation}
When the chemical potential exceeds the critical value, $\mu>\mu_{\rm c}$, 
the saddle point equation breaks down and the average $r=r(\mu)$  drops to zero which, remembering $r$ is an inverse density,  is the grand-canonical version of the condensation transition in the model.
For power-law weights,  the critical value of the chemical potential 
is $\mu_{\rm c}=\ln \zeta(\beta)$.  

The singularities of grand-canonical potential
$\psi(\mu)$ for power-law weights 
at the critical chemical potential $\mu_{\rm c}=\ln \zeta(\beta)$ for $\beta>1$
can be determined.
The critical behaviour corresponds to $\mu_{\rm c} - \mu \rightarrow 0^+$. 
It is found that the phase transition is first order for $\beta\in [2,+\infty)$ and higher order for $\beta\in (1,2)$ \cite{balls2,bbj2}. 
More precisely, for $\beta \in (1,2)$ and
for $\mu_{\rm c}-\mu \rightarrow 0^+$
\begin{equation}
    \psi''(\mu) \sim (\mu_{\rm c} -\mu)^{-\lambda} ,
    \label{eq:div}
\end{equation}
where 
\begin{equation}
    \lambda = 2 - \frac{1}{\beta-1} .
    \label{eq:l2}
\end{equation}
The symbol $\sim$ stands for the most singular part for
$\mu_{\rm c}-\mu \rightarrow 0^+$. We see that $\lambda\in [0,1)$ for $\beta \in [3/2,2)$, so the transition is second
order in this range of $\beta$. More generally, it can be shown
\cite{balls2} that the transition is of $n$-th order for $\beta \in [1+1/n,1 + 1/(n-1))$. 
The transition disappears at $\beta=1$ as the critical value of $\mu$ is pushed to infinity. 

The zeta-urn model in the grand-canonical ensemble thus provides a model which 
displays both first order and continuous transitions of arbitrary order, as $\beta$ is varied and whose partition function may be explicitly calculated for finite size systems by  virtue of (\ref{eq:Z-recurrence})  and  (\ref{ZSmu}). In the remainder of the paper we  exploit this to use it as a testing ground for examining the scaling properties of partition function zeros for transitions of various orders. Before doing this we set the scene with a discussion of the general properties of partition function zeros and their finite size scaling.

\section{Partition function zeros}
\label{sec:PFZ}

Regarding the partition function for a finite size system as a polynomial in some fugacity gives an appealing picture for understanding how singularities appear in the thermodynamic limit. For external field driven transitions, Lee and Yang  \cite{LY1,LY2} made the key observation that if one considered  complex fugacities,  physical transitions appeared as the complex zeros of the partition function polynomial moved in to pinch the real positive axis in the complex fugacity plane in the thermodynamic limit. This point of view was then extended by Fisher to temperature driven transitions \cite{Fi64}.
With a view to the application of this observation to the grand-canonical zeta-urn model discussed previously,
we write the partition function for a system of size $S$, $Z_S$, as
a polynomial in the fugacity, $z=\exp(-\mu)$ 
\begin{equation}
 Z_S(z) = \prod_{j=1}^S{\left( z-z_j(S) \right)}
 ,
\label{Z}
\end{equation}
where $j$ labels the zeros. Clearly, the two forms of the partition function (\ref{ZSmu}) and (\ref{Z}) are equivalent 
$Z_{S,\mu} = Z_S(\re^{-\mu})$. One can define
a thermodynamic potential
\begin{equation}
 f_S(z) = \frac{1}{S} \ln{Z_S(z)}
 =
 \frac{1}{S} \sum_j{\ln{\left( z - z_j(S) \right)}}
 ,
\label{f}
\end{equation}
Taking the limit as $S \to \infty$ leads to 
\begin{equation}
    f_\infty(z) = \int \rd^2\xi \varrho(\xi) \ln (z-\xi) ,
    \label{frho}
\end{equation}
where $\varrho(\xi)$ is the density of zeros on the complex plane.
As $\ln|z-\xi|/2\piup$ is the Green function for the 2D Poisson
equation, the last equation has an electrostatic analogy \cite{BDL}, 
with $\varrho(\xi)$ interpreted as charge density, 
and $\Re f_\infty(z)$ as electrostatic potential. 
For many statistical systems that undergo a phase
transition, the distribution
of zeros is known to display a very specific pattern.
Near the critical point $x_{\rm c} = \re^{-\mu_{\rm c}}$ which is located on the positive real semi-axis, 
zeros lie on a one dimensional curve. This curve is symmetric with respect to the real axis because the partition function has real coefficients and as 
a consequence non-real zeros come in conjugate pairs. 
In the limit $S\rightarrow \infty$, the density of zeros on this curve
near the critical point can be parameterized by a single parameter, which we take to be the imaginary part of zeros.
For simplicity assume that the positive branch of this curve
hits the real axis with a non-zero angle $\gamma$. 
Near the critical point the positive branch can be approximated by $\xi(y) = x_{\rm c} + c y + i y$ for $y < \epsilon \ll 1$, where
$c = \cot(\gamma)$. The range $\epsilon$ is chosen as small as possible, so that 
higher powers of $y$ in $\xi(y)$ can be neglected. For $\gamma=0$ or $\piup$, 
that is when the curve is tangent to the real axis, the parameterisation 
of this curve by $y$ is not optimal.
However, this is not the 
case for the curves discussed below.

For a real argument $z=x$ (\ref{frho}) takes on the form  
\begin{equation}
    f_\infty(x) = \int_0^{\epsilon} 
    \rd y g(y) \ln \left[(\Delta x - cy)^2 + y^2\right] + \ldots,
\end{equation}
where $\Delta x = x-x_{\rm c}$ and $g(y)$ is the density of zeros on the line 
$\xi(y) = x_{\rm c} + c y + \ri y $
(for $y < \epsilon \ll 1$). The dots denote the contribution from the remaining zeros, lying further away from the critical point.
Taking the derivative of both sides and changing the integration
variable from $y$ to $w=y/|\Delta x|$, we obtain
\begin{equation}
    f'_\infty(x) = \int_0^{\epsilon/|\Delta x|} 
    \rd w g\left(w|\Delta x|\right) 
    \frac{2(-cw \pm 1)}{(cw\pm 1)^2 + w^2}
     + \ldots,
     \label{fprim}
\end{equation}
where the signs $\pm$ correspond to $x = x_{\rm c} \pm |\Delta x|$.
When the density of zeros on the curve 
approaches a finite constant 
$g(y) \rightarrow g_0 > 0$ for $y\rightarrow 0$, 
then the integrals (\ref{fprim}) on both sides of $x_{\rm c}$ may take on different values, leading to a discontinuity of $f'_\infty(x)$ at $x=x_{\rm c}$, and a first order phase transition.  One can show that in this case zeros approach the critical point
perpendicularly to the real axis \cite{GR}, meaning that the coefficient $c$
in equation~(\ref{fprim}) equals zero. If $g(y) \rightarrow 0$ for $y\rightarrow 0$, then the discontinuity of the first derivative
disappears, but higher derivatives may have a discontinuity or a
divergence at the critical point. In particular, when the density
behaves as $g(y) \sim y^{1-p}$ for $y\rightarrow 0$, 
where $p \in (0,1)$ then the  contribution from the integral
in (\ref{fprim}) tends to zero as $f'_\infty(x) \sim |\Delta x|^{1-p}$, 
but the next derivative diverges
\begin{equation}
    f''_\infty(x) \sim |\Delta x|^{-p} ,
\end{equation}
so the transition is second order. The grand-canonical thermodynamic potential $\psi(\mu)$ is directly related to $f_\infty(x)$:
$\psi(\mu) = f_\infty(\re^{-\mu})$. This means, in particular,
that the divergence of $f''_\infty(x)$ for $x \rightarrow x_{\rm c}^+$ 
leads to a divergence of $\psi''(\mu)$ for 
$\mu \rightarrow \mu_{\rm c}^-$ with the same exponent $p$
\begin{equation}
    \psi''(\mu) \sim (\mu_c - \mu)^{-p} .
    \label{eq:psi2}
\end{equation}
Comparing the last equation with (\ref{eq:div}), we see that for the zeta-urn model
\begin{equation}
p=\lambda = 2- \frac{1}{\beta-1} .
\label{eq:pbeta}
\end{equation}
The exponent describing the divergence of the second derivative 
of the grand-canonical thermodynamic potential is thus
directly related to the exponent describing the behaviour of the 
density of zeros on the curve near the critical point ($y\rightarrow 0$)
\begin{equation}
    g(y) \sim y^{1-p} ,
\end{equation}
or equivalently to the exponent of the cumulative density function
$G(y) = \int_0^y g(y') \rd y'$
\begin{equation}
    G(y) \sim y^{2-p} .
    \label{eq:Glimit}
\end{equation}
The cumulative density $G(y)$ can be reconstructed from its counterpart $G_S(y)$ for finite $S$, by sending $S$ to infinity $G(z) = G_\infty(z) \equiv \lim_{S\rightarrow \infty} G_S(z)$.
Let $z_j = x_j + \ri y_j$ be the loci of zeros for finite $S$. The index $j$ goes from $1$ to $S$.
The zeros $z_j$ in the upper half plane are indexed in order of increasing distance 
$|z_j-x_{\rm c}|$ from the critical point. 
If the first zeros indeed lie on a smooth curve, 
then  the imaginary parts are also ordered $y_1\leqslant y_2 \leqslant y_3 \ldots$, $j=1,2,\ldots,$. 
The conjugate zeros which are located symmetrically in the lower plane can also be ordered, 
for example $z_{S+1-j} = \bar{z}_j = x_j - \ri y_j$, 
but the way they are ordered is inessential 
for the calculation which concentrates on the 
first zero in the upper half plane. The cumulative density for the first zeros is:
\begin{equation}
    G_S(y_j) = \frac{j}{S} ,
    \label{eq:GS}
\end{equation}
because the right-hand side gives the fraction of all zeros which lie near the critical point 
and whose imaginary part $y \in [0,y_j]$. Generally, $G_S(y)$ is a staircase function with steps of height
$1/S$ which decrease to zero when $S$ tends to infinity. For large $S$, the function $G_S(y)$ should approach the limiting distribution (\ref{eq:Glimit}), so for large $S$ we can expect that the zeros must scale as 
\begin{equation}
    y_j \sim \left(\frac{j}{S}\right)^{\frac{1}{2-p}},
    \label{eq:yj}
\end{equation}
in order to correctly reproduce the right-hand side of (\ref{eq:GS}) for large $S$. In other words, we see that the phase transition properties are encoded in the scaling of the loci of zeros closest to the critical point. In particular, we see that for 
$y_j \sim j$ for $p=1$, so the zeros are equidistant. For $p=0$, which is a
marginal value for the transitions of second and third order, the last formula gives 
$y_j \sim j^{1/2}$. For the zeta-urn model,
equations (\ref{eq:Glimit}) and (\ref{eq:yj})
translate to 
\begin{equation}
    G(y) \sim y \  , \  \ \ y_j \sim \frac{j}{S},
\end{equation}
for $\beta \in [2,+\infty)$, where the transition is
first order, and 
\begin{equation}
    G(y) \sim y^{\frac{1}{\beta-1}} \ , \ \ \  y_j \sim \left(\frac{j}{S}\right)^{\beta-1},
    \label{eq:Gy}
\end{equation}
for $\beta \in [3/2,2)$ (\ref{eq:pbeta}), 
where the transition is
second order. What happens for $\beta \in (1,3/2)$,
where the transition is higher order, requires a further
discussion. The expression on the left-hand
side of (\ref{eq:psi2}) corresponds to normalised variance of $N$ 
\begin{equation}
    \psi''(\mu) 
      =  \lim_{S\rightarrow \infty}
    \frac{\langle N^2 \rangle_{S,\mu} - \langle N \rangle_{S,\mu}^2}{S}
    = \lim_{S\rightarrow \infty}
    \frac{\sigma^2_{S,\mu}(N)}{S}
    \label{eq:psibis_limit}
\end{equation}
as directly follows from the definition 
(\ref{eq:psi}). In the fluid phase, for a given $\mu < \mu_{\rm c}$, 
the fluctuations of $N$
grow as a square root of $S$: $\sigma_{S,\mu}(N) \sim \sqrt{S}$, so the  
normalised variance tends to a constant, $\psi''(\mu)$, as
$S$ tends to infinity. For the first and second order phase
transitions, at the critical point $\mu \rightarrow \mu_{\rm c}$, the fluctuations grow faster than $\sqrt{S}$ leading to the divergence of $\psi''(\mu)$ in the limit $S\rightarrow \infty$ 
(\ref{eq:psibis_limit}). For the third or higher order phase transition,
however, the fluctuations at the critical point behave asymptotically as 
$\sigma_{S,\mu}(N) = \sigma_{\rm c} \sqrt{S} + o(\sqrt{S})$. If the
coefficient $\sigma_{\rm c}$ is non-zero, then the normalised 
variance (\ref{eq:psibis_limit}) tends to a positive constant 
$\psi''(\mu_{\rm c}) = \sigma_{\rm c}>0$ at the critical point. In this case, $p=0$, as can be seen from~(\ref{eq:psi2}).
If in turn the leading term vanishes, $\sigma_{\rm c}=0$, then
$\psi''(\mu_{\rm c})$ tends to zero and as a consequence
the exponent $p$ (\ref{eq:psi2}) takes on a negative value, which is the case for the zeta-urn model. More generally, one can show 
\cite{balls2} that all derivatives $\psi^{(k)}(\mu_{\rm c})=0$, for $k=1,\ldots,n$, 
vanish at the critical point for $\beta \in [1+1/(n+1),1+1/n)$, and, as a consequence,  
the scaling relations (\ref{eq:Gy}) apply to the entire range $\beta \in (1,2)$, where the phase 
transition is second or higher order.

\section{Finite size analysis}
\label{sec:FSS}

For moderate system sizes, up to $S$ of the order a thousand, 
computations of the partition function for the zeta-urn 
model (\ref{eq:ZSN}) can be done directly by algebraic calculations.
The procedure is straightforward. We first specify
$\beta$, hence fixing the order of the transition, 
and $S$, and then evaluate $Z_{S,N}$'s 
for the requisite $N=1,\ldots,S$ using the 
recursion relation of (\ref{eq:Z-recurrence}). 
The roots of the polynomial
\begin{equation}
   Z_S(z) = \sum_{N=1}^S z^N Z_{S,N},
\end{equation}
are then extracted to obtain the partition function zeros in the complex $z=\exp(-\mu)$ plane.
We performed computations for sizes up to $S=2000$, using Mathematica$^{\circledR}$.

To estimate $G(y)$, we select the zeros
nearest to the critical point in the upper half-plane
and fit the function 
\begin{equation}
G(y)=ay^b+c,
\label{eq:Gabc}
\end{equation}
with three parameters $a,b,c$ to the 
data points $(y_j,(j-1/2)/S)$, $j=1,\ldots,n$ (\ref{eq:GS}) for the first $n$ zeros 
\cite{wjk}. 
The parameter $c$ is a finite size correction, which is expected to tend to zero when $S$ tends to infinity. From the exponent $b$ we can estimate $p=2-b$. For first order phase transitions $b=1$,
while for higher order transitions $b=1/(\beta-1)$. 

We can also estimate the exponent directly from the positions
of the first zeros (\ref{eq:yj}). In the analysis we should take into
account possible finite size corrections. The simplest finite size
corrections to (\ref{eq:yj}) have the form
\begin{equation}
    y_j = \left(\frac{j+c}{aS+b}\right)^d,
    \label{eq:yfs}
\end{equation}
where $d=1/(2-p)$, and $a,b,c$ are real parameters representing
finite size corrections to (\ref{eq:yj}). We have used 
the same letters here as in (\ref{eq:Gabc}) but now they have 
a different meaning.
This formula can be applied with a free parameter $d$ 
which can be fitted from the data and then can be used to estimate the
exponent~$p$. Alternatively, the value of $d$ can be fixed  
theoretically, to check whether the theoretical 
prediction describes the data well. In this case, only $a,b$ and $c$
are fitted. We will take the latter approach here,
setting $d=1$ 
for $\beta\in [2,+\infty)$ and $d=\beta-1$ for $\beta \in (1,2)$ according to the theoretical predictions.

\section{Results}
\label{sec:RES}
\subsection{First order regime}
\label{first}

\begin{figure}[h!]
    \centering
    \includegraphics[width=0.4\textwidth]{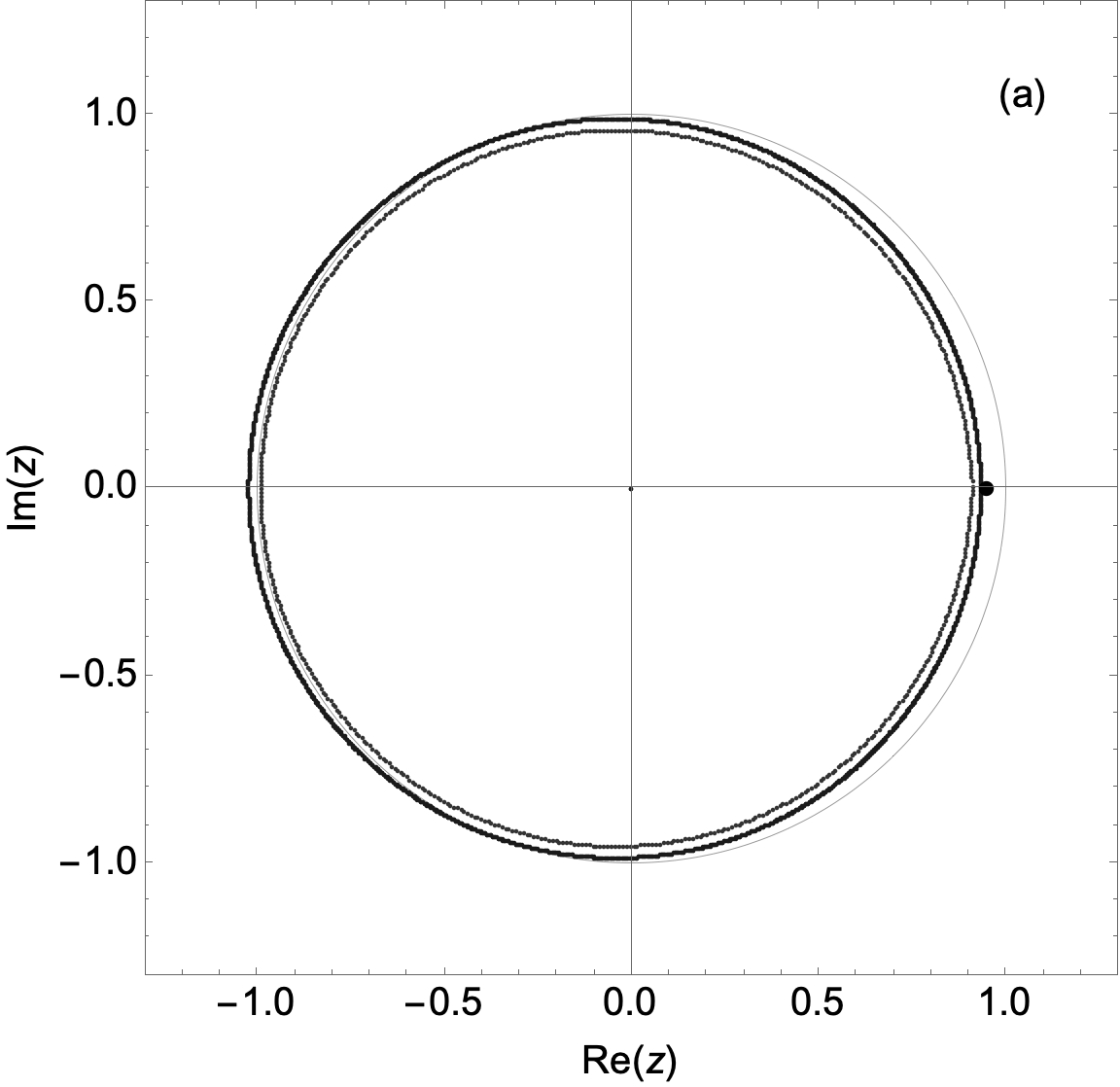} 
\qquad 
    \includegraphics[width=0.5\textwidth]{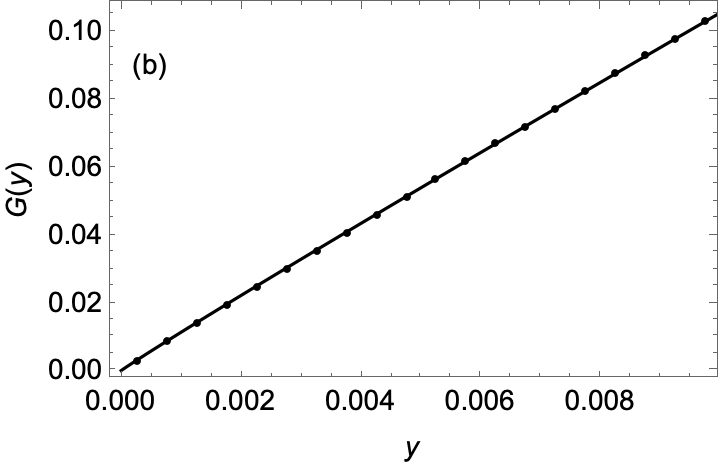} 
    \caption{(a) Partition function zeros for $\beta=9/2$ for $S=600$ (gray) and $S=2000$
    (black). One can see that the contour on which the zeros lie expands slightly with $S$. The contours are close to the unit circle which is drawn as thin line. The black point on the real axis
    at $\exp(-\mu_c)= 1/\zeta(9/2) \simeq 0.94813$ is drawn to highlight that the zeros are impacting the real axis vertically close to the expected (infinite size) critical value.
    (b) Fitting the cumulative density for the first $20$ zeros for $\beta=9/2$ and $S=2000$ 
    to $G(y) = a y^b + c$ gives $a=0.1687(6)$, $b=1.028(1)$ and $c=-0.7(5) \cdot 10^{-5}$.
    }
    \label{fig:b92}
\end{figure}

As an example we take $\beta=9/2$, where the transition is well into the first order regime ($\beta > 2$). The expectation is that the zeros will impact the positive real axis in the complex fugacity plane at the pseudo-critical point $z_{\rm pc}(S)$, which will tend to the critical point $z_{\rm c}(S) = \exp(-\mu_{\rm c}) = 1/\zeta(\beta)$ for $S\rightarrow \infty$. For $\beta=9/2$, $z_{\rm c} =1/\zeta(9/2) \approx 0.94813$. The change of the position of the pseudo-critical point with $S$ can be seen in 
figure~\ref{fig:b92}(a) where the zeros are plotted for $S=600$ and $S=2000$.  
In figure~\ref{fig:b92}(b) we show the best
fit $G(y)=ay^b+c$ (\ref{eq:Gabc}) to $n=20$ first zeros. The best fit gives the power $b=1.032$ and $b=1.028$ for $S=600$ and $S=2000$, respectively,
exhibiting a tendency to move towards 
$b=1$, expected for a first order transition. The fitted value of $c$ (\ref{eq:Gabc}) is $c=-1.6 \cdot 10^{-5}$ 
and $-0.7 \cdot 10^{-5}$ for $S=600$ and $S=2000$, so it tends to zero with increasing $S$, as expected. 
\begin{figure}[h!]
    \centering
    \includegraphics[width=0.45\textwidth]{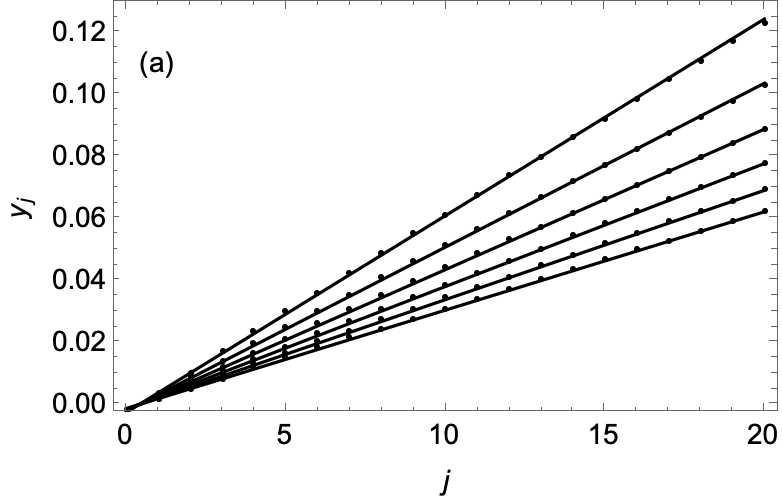} 
\qquad
    \includegraphics[width=0.45\textwidth]{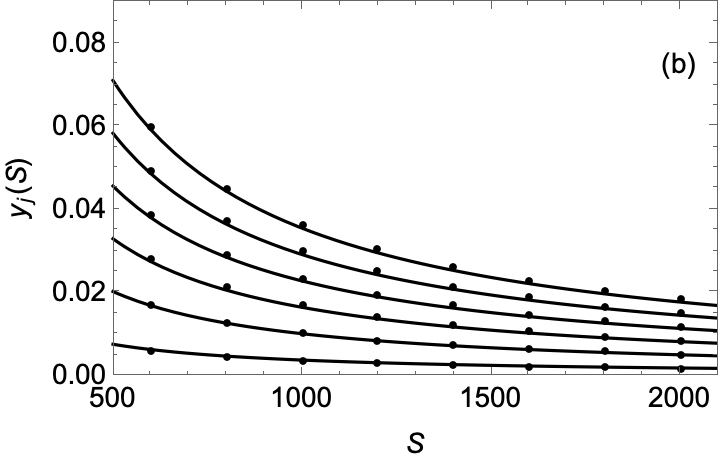} 
    \caption{(a) The imaginary parts $y_j$ of the first twenty zeros plotted against $j$, for 
    $S=1000, 1200, \ldots, 2000$ from top to bottom. The solid lines are given by  (\ref{eq:yfs}) with $a=0.1575$, $b=0$,
    $c=5.64$ and $d=1$. (b) The imaginary parts 
    $y_j(S)$ of the first six zeros plotted against $S$ for $j=1,2,\ldots, 6$ from bottom to top.
    The solid lines are given by (\ref{eq:yfs}) with the same parameters as in (a).}
    \label{fig:yj_b92}
\end{figure}
An alternative check of the scaling of first zeros, expected
for the first order phase transition, is to
use (\ref{eq:yfs}) with $d=1$. As shown in
figure~\ref{fig:yj_b92}) the
scaling formula (\ref{eq:yfs}) very well reproduces the dependence $y_j(S)$ both
as function of $j$ and $S$.

Another property that we can check from the distribution of the zeros in the first order case is magnitude of the discontinuity in $\psi'(\mu)$ at $\mu_{\rm c}$, which is given theoretically from the saddle point solution as $\Delta \psi' (\mu_{\rm c})= \zeta(\beta)/\zeta(\beta-1)$ \cite{balls2}. For $\beta=9/2$, the numerical value of this discontinuity is $\Delta \psi' (\mu_{\rm c}) \approx 0.93608$. On the other hand, this value is related to the position of the first zero \cite{LY1,LY2}, or equivalently to the parameter $a$ (\ref{eq:Gabc}) as 
$\Delta \psi'(\mu_{\rm c}) \approx 2 \piup a$.
The numerical value of $2 \piup a$ deduced from the partition function zeros for $S=2000$ is 
$2 \piup \times 0.1687 \approx 1.059 $ which is
in a moderately good agreement with the saddle point value. One would expect a better agreement for larger $S$.

\subsection{Second (and higher) order regime}
\label{second}

As $\beta \to 2$ and the strength of the first order transition weakens and the distribution becomes less circular.
\begin{figure}[h!]
     \centering
    \includegraphics[width=0.4\textwidth]{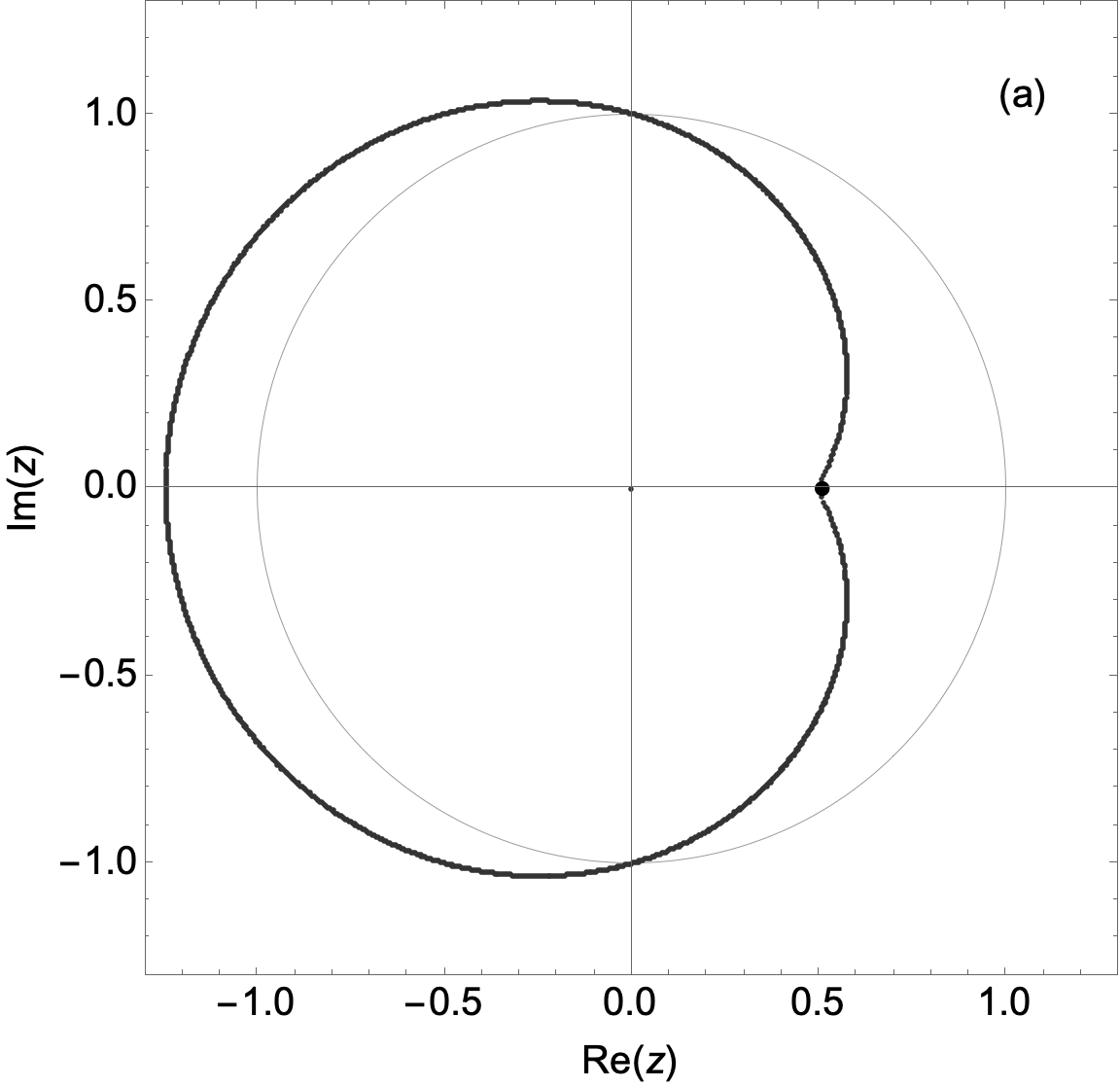} \qquad 
    \includegraphics[width=0.5\textwidth]{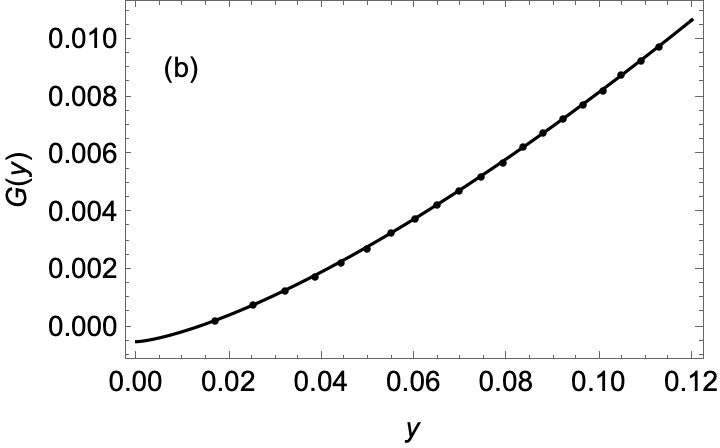} 
    \caption{(a) Partition function zeros for $\beta=7/4$ and $S=1200$. The unit circle as reference is drawn with a thin line. The critical point is drawn as a black point on the real axis at $\exp(-\mu_c)= 1/\zeta(7/4)=0.50960$. The line going through the smallest zeros hits the real axis at a pseudo-critical point which is slightly away from the critical point. The pseudo-critical point moves towards the critical point as $S$ is increased.
    (b) Fitting the cumulative density for the first $20$ zeros for $\beta=7/4$ and $S=2000$ 
    to $G(y) = a y^b + c$ gives $a=0.213(2)$, $b=1.390(6)$ and $c=-0.52(2) \cdot 10^{-3}$.
    }
    \label{fig:b74}
\end{figure}
Since $\psi''(\mu)$ displays a logarithmic singularity at $\beta=2$   it is not a particularly friendly value for numerical explorations.  Out of an abundance of caution we move into the range of $\beta$ where the transition is second order and does {\it not} have logarithmic corrections, taking $\beta=7/4$, which we show in  figure~\ref{fig:b74}. 
From (\ref{eq:div}) this gives a diverging second order phase transition with
$\psi''(\mu) \sim (\mu_{\rm c} - \mu)^{-2/3}$ (\ref{eq:l2}) and $G(y) \sim y^{4/3}$ (\ref{eq:Gy}). The best fit (\ref{eq:Gabc})
produces a value $b=1.39$ close to the expected 
one. The fit is shown in figure~\ref{fig:b74}(b).
\begin{figure}[h!]
    \centering
    \includegraphics[width=0.45\textwidth]{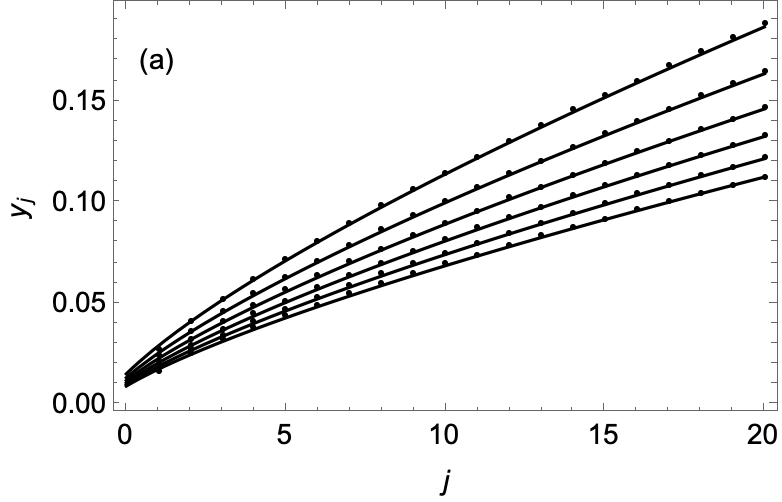} \qquad
    \includegraphics[width=0.45\textwidth]{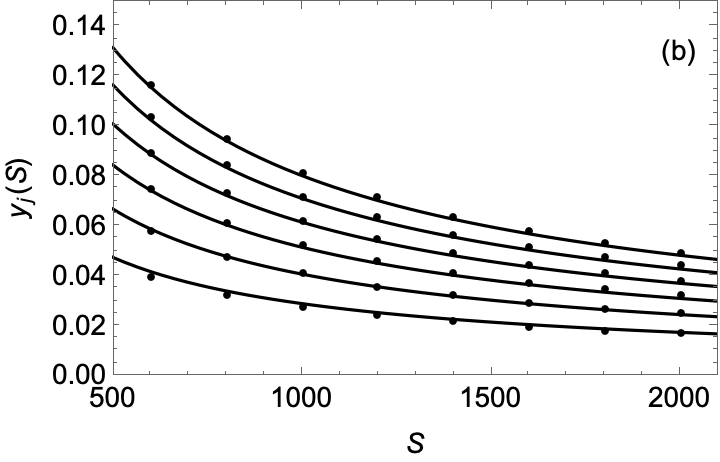} 
    \caption{(a) The imaginary parts $y_j$ of the first twenty zeros plotted against $j$, for 
    $S=1000, 1200, \ldots, 2000$ from top to bottom. The solid lines are given by  (\ref{eq:yfs}) with $a=0.73$, $b=0.1875$,
    $c=6.7$ and $d=3/4$. (b) The imaginary parts 
    $y_j(S)$ of the first six zeros plotted against $S$ for $j=1,2,\ldots, 6$ from bottom to top.
    The solid lines are given by (\ref{eq:yfs}) with the same parameters as in (a).}
    \label{fig:yj_b74}
\end{figure}
Also the dependence of $y_j(S)$ on $j$ and
$S$ is consistent with (\ref{eq:yfs}) using the
power $d=\beta-1=3/4$ as shown in figure~\ref{fig:yj_b74}.

The transition will be $n$-th order for 
$\beta \in [1 + 1/n, 1/(n-1))$. 
The general form of the locus of zeros does not change as $\beta \to 1$, though the pinch point on the real axis tends to the origin, $z_{\rm c} \to 0$, as $\mu_{\rm c} \to \infty$. We will present a quantitative investigation of the higher order scaling elsewhere. As mentioned, the 
\emph{ansatz} $G(y) \sim y^{1/(\beta-1)}$ appears to extend to the higher order transition region for
the entire range $\beta \in (1,2)$ but for
$\beta \in (1,3/2]$ it is  difficult to discern using a finite size scaling analysis because it is overshadowed by the behaviour $G(y) \sim y^2$ coming from the maximum of the second derivative, which lies very close to the critical point.

\subsection{No transition regime}
\label{unphysical}

\begin{figure}[h!]
    \centering
    \includegraphics[width=0.45\textwidth]{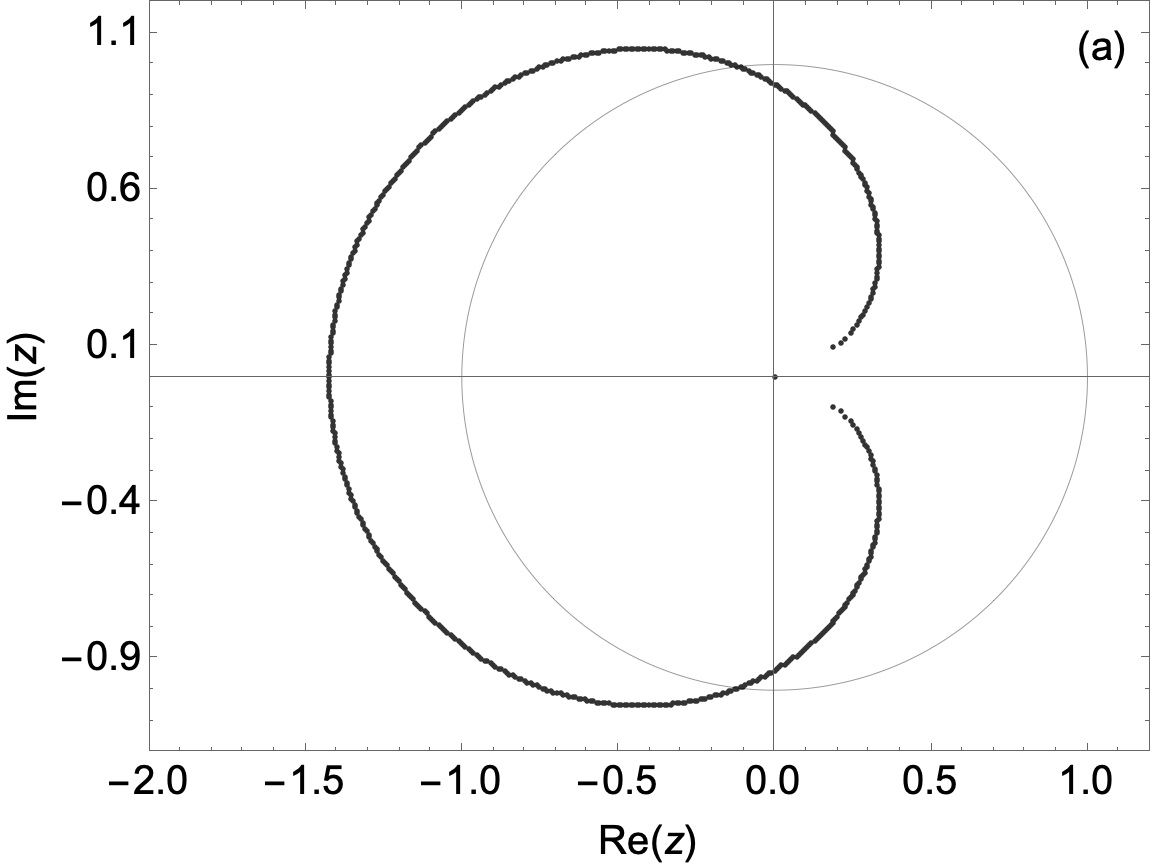} 
    \includegraphics[width=0.45\textwidth]{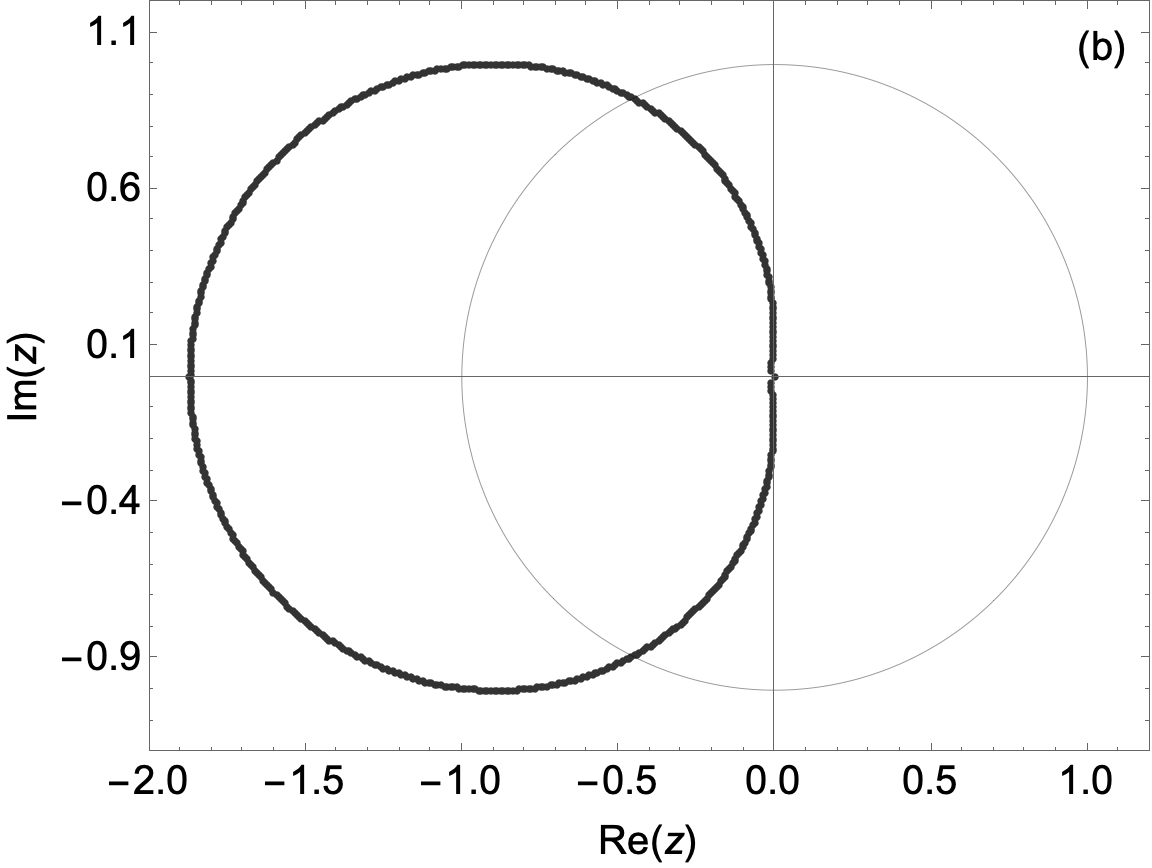}
    \qquad
    \caption{(a) Partition function zeros for $\beta=1$ and $S=600$. (b)
    Partition function zeros 
    for $\beta=1/8$ and $S=600$. The unit circle shown for reference is drawn with a thin line.}
    \label{fig:beta-lt1}
\end{figure}

Although the transition vanishes at $\beta=1$, it is still possible to evaluate the zeros numerically for $0\leqslant \beta \leqslant 1$ and, indeed, for $\beta<0$. As expected, the locus of zeros does not impact the (physical) positive real $z$ axis but, nonetheless, still displays interesting behaviour. If we first consider $\beta=1$, we can see the locus of zeros failing to pinch the real axis in figure \ref{fig:beta-lt1} (left-hand). As $\beta$ is reduced further, the locus of zeros moves into the negative half plane as can be seen in figure \ref{fig:beta-lt1} (right-hand) for $\beta=1/8$.

\begin{figure}[h!]
    \centering
    \includegraphics[width=0.45\textwidth]{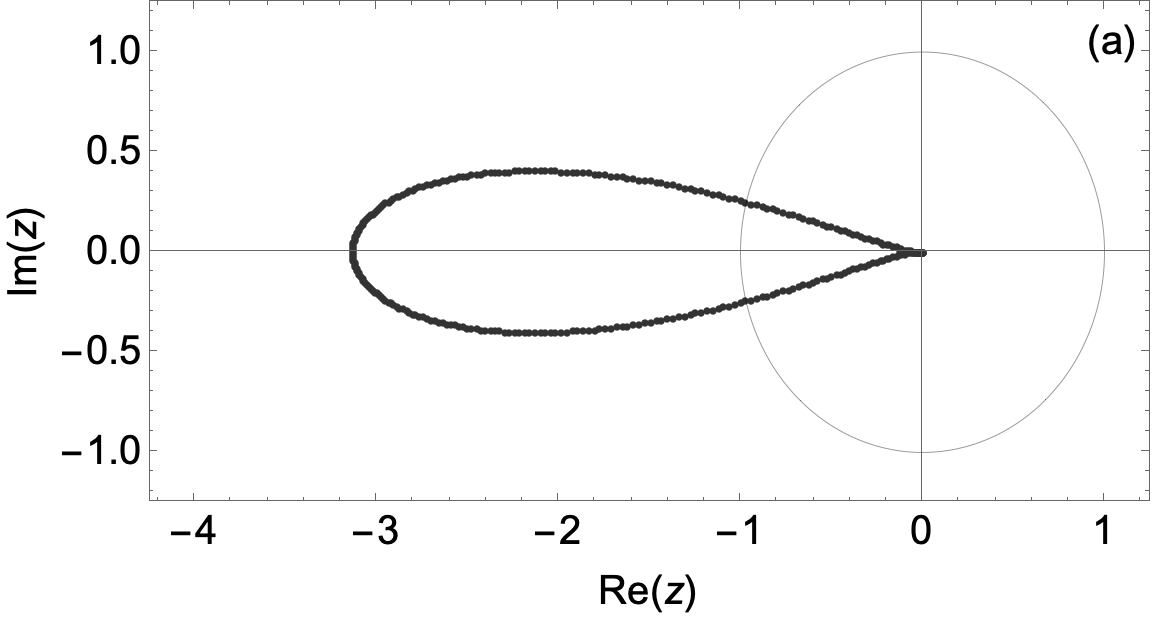} 
    \includegraphics[width=0.45\textwidth]{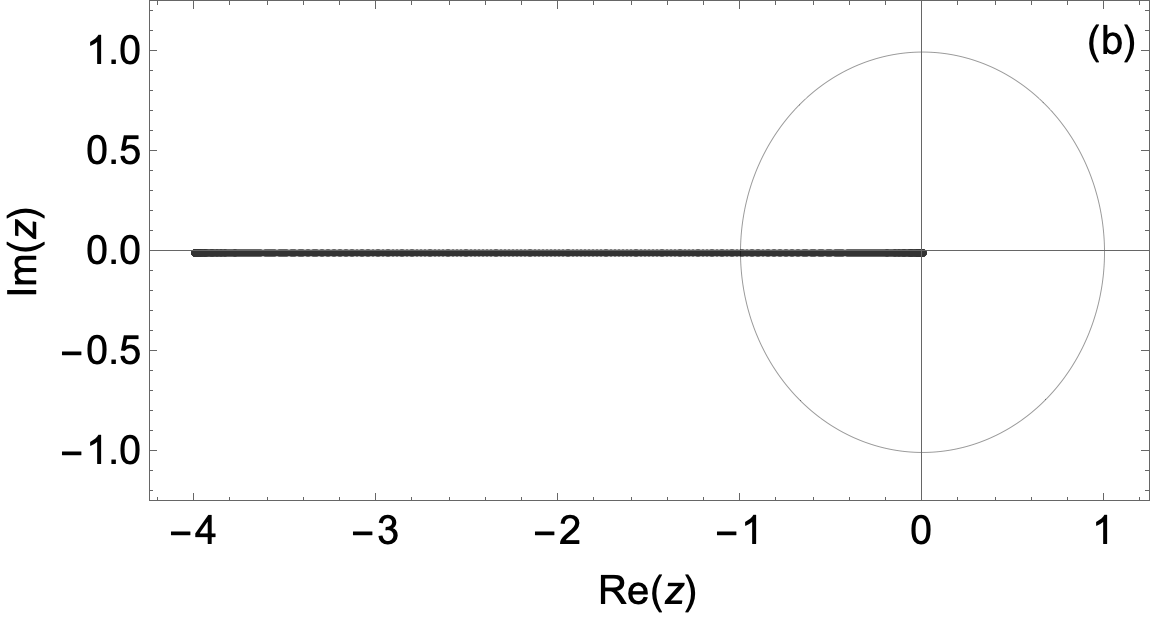}
    \qquad
    \caption{(a) Partition function zeros for $\beta=-3/4$ and $S=300$. (b) Partition function zeros 
    for $\beta=-1$ and $S=300$. The unit circle shown for reference is drawn with a thin line.}
    \label{fig:beta-minus}
\end{figure}

It is possible to evaluate $Z_{S,\mu}$ explicitly for $\beta=0$ which was done in \cite{balls2}, where it was observed that 
$Z_{S,N} = \binom{S-1}{N-1}$
and the sum in~(\ref{ZSmu}) was then carried out to  
give $Z_{S,\mu} = z(1+z)^{S-1}$ (with $z=\exp(-\mu)$). In this case, the locus of zeros degenerates to the single trivial zero at the origin and an $(S-1)$-fold zero at $z=-1$. 
For $-1<\beta<0$, the locus takes on an airfoil-like shape which thins and extends further left as $\beta \to -1$, eventually collapsing onto the negative real axis for $\beta \leqslant -1$ as can be seen in 
figure \ref{fig:beta-minus}.
\

\newpage
\section{Summary}
\label{sec:summary}

The analysis of the partition function zeros of the zeta-urn model carried out here  has some parallels with that done for the ASEP (asymmetric exclusion process) in \cite{be1,be2}. In that case the first and second order nature of the transitions in an ostensibly non-equilibrium model were discernable from ``partition function'' zeros which were explicitly calculable from the known exact formula for the analogue of the partition function (a steady state normalization).  It was later observed that the weights of ASEP configurations could also be regarded as those of an equilibrium model of one transit walks, so the ASEP steady state normalization {\it was} a standard partition function for this model \cite{bjjk1,bjjk2}. Here, the zeta-urn partition function can also be regarded as the non-equilibrium steady state of a zero range process with suitably tuned jump rates. One obvious difference to the ASEP is that transitions of any order are accessible for the zeta-urn here by tuning $\beta$, another is that the ensemble considered in the ASEP case was the canonical one. 

We have found that the zeta-urn model provides a useful pedagogical illustration of the finite size scaling of partition function zeros for both first order ($\beta>2$) and continuous ($1< \beta \leqslant 2$) transitions, by virtue of the relative ease with which the finite size partition functions may be calculated, thanks to~(\ref{eq:Z-recurrence}) and (\ref{ZSmu}) . Moving away from  statistical mechanics, the locus of zeros itself continues to display interesting behaviour for regions in which there is no phase transition (i.e. regions in which the locus does not intersect the positive real axis in the $S \to \infty$ limit). Given that $\psi(\mu)$ is the inverse of a known cumulant generating function,  it is tempting to try to find analytical expressions for these curves.  

It would also be interesting to devote further numerical resources to exploring the logarithmic corrections at $\beta=2$ and to use the model to benchmark alternative approaches such as the use of high-order cumulants to extract the scaling of the zeros \cite{DBF,VLF}.

\section*{Acknowledgments}

Ralph Kenna contributed greatly to DAJ's understanding of partition function zeros and finite size scaling, through joint work such as \cite{wjk,bjjk2} and elsewhere. He will be sadly missed.

\ukrainianpart

\title{Нулі статистичної суми моделі дзета-урн}
\author{П. Б'ялас\refaddr{label1},  З. Бурда\refaddr{label2}, Д. А. Джонстон\refaddr{label3}}
\addresses{
	\addr{label1} Інститут прикладної інформатики Ягеллонського університету, вул. Лоясієвіча 11, 30-348 Краків, Польща
	\addr{label2}Гірничо-металургійна академія, Факультет фізики та прикладної інформатики, 
	алея Міцкевича 30, 30-059 Краків, Польща
	\addr{label3} Школа математики та комп’ютерних наук, Університет Геріот-Ватт, Ріккартон, Единбург EH14 4AS, Великобританія
}

\makeukrtitle

\begin{abstract}
	\tolerance=3000%
Обговорюється розподіл нулів статистичної суми для великого канонічного ансамблю у моделі дзета-урн, де налаштування одного параметра може дати конденсаційний перехід першого чи будь-якого вищого порядку.  Обчислюється геометричне місце нулів для систем скінченого розміру та перевіряються масштабні співвідношення, що описують накопичення нулів поблизу критичної точки у порівнянні з теоретичними передбаченнями як для перехідних режимів першого, так і вищих порядків.
	\keywords Нулі Лі-Янга та Фішера, критичні показники, фазові переходи першого та другого роду
	
\end{abstract}
\end{document}